\documentclass[twocolumn,10pt]{IEEEtran}
\usepackage{ifpdf, flushend,subfigure}

\ifCLASSINFOpdf
  \usepackage[pdftex]{graphicx}
  \graphicspath{{../pdf/}{../jpeg/}}
  \DeclareGraphicsExtensions{.pdf,.jpeg,.png}
\else
  \usepackage[dvips]{graphicx}
  \graphicspath{{../eps/}}
  \DeclareGraphicsExtensions{.eps}
\fi
\usepackage[cmex10]{amsmath}
\usepackage {amssymb}
\usepackage{algorithmic}
\usepackage{array}
\usepackage{mdwmath}
\usepackage{mdwtab}
\usepackage{eqparbox}
\usepackage{url}
\usepackage{algorithm}
\usepackage{algorithmic}

\newcommand{\argmin}{\operatornamewithlimits{argmin}}
\newcommand{\argmax}{\operatornamewithlimits{argmax}}
\newcommand{\beq}{\begin{equation}}
\newcommand{\eeq}{\end{equation}}
\newcommand{\beqn}{\begin{eqnarray}}
\newcommand{\eeqn}{\end{eqnarray}}
\newcommand{\beqno}{\begin{eqnarray*}}
\newcommand{\eeqno}{\end{eqnarray*}}
\newcommand{\bma}{\begin{displaymath}}
\newcommand{\ema}{\end{displaymath}}
\newcommand{\bnu}{\begin{enumerate}}
\newcommand{\enu}{\end{enumerate}}
\newcommand{\bce}{\begin{center}}
\newcommand{\ece}{\end{center}}
\newcommand{\btb}{\begin{tabular}}
\newcommand{\etb}{\end{tabular}}

\begin{document}
%
\title{Fair Channel Allocation and Access Design for Cognitive Ad Hoc Networks}

\author{\IEEEauthorblockN{Le Thanh Tan and Long Bao Le}  
\thanks{The authors are with INRS-EMT, University of Quebec,  Montr\'{e}al, Qu\'{e}bec, Canada. 
Emails: \{lethanh,long.le\}@emt.inrs.ca. }}

\maketitle

\begin{abstract}
We investigate the fair channel assignment and access design problem for 
cognitive radio ad hoc network in this paper. In particular, we consider
a scenario where ad hoc network nodes have hardware constraints which allow
them to access at most one channel at any time. We investigate a fair channel allocation
 problem where each node is allocated a subset of channels which are sensed and accessed periodically by their owners
 by using a MAC protocol.
Toward this end, we analyze the complexity of the optimal brute-force search algorithm
which finds the optimal solution for this NP-hard problem. We then develop low-complexity
algorithms that can work efficiently with a MAC protocol algorithm, which resolves the
access contention from neighboring secondary nodes. Also, we develop a throughput
analytical model, which is used in the proposed channel allocation algorithm and for
performance evaluation of its performance. Finally, we present extensive numerical
results to demonstrate the efficacy of the proposed algorithms in achieving fair
spectrum sharing among traffic flows in the network.
\end{abstract}


\section{Introduction}

Cognitive radio has recently emerged as an important research field, which promises
to fundamentally enhance wireless network capacity in future wireless system.
To exploit spectrum opportunities on a given set of channels of interest,
each cognitive radio node must typically rely on spectrum sensing and access mechanisms.
In particular, an efficient spectrum sensing scheme aims at discovering spectrum holes
in a timely and accurate manner while a spectrum access strategy coordinates the spectrum
access of different cognitive nodes so that high spectrum utilization can be achieved.
These research themes have been extensively investigated by many researchers in recent years
\cite{Yu09}-\cite{tan12}. In \cite{Yu09}, a survey of recent advances in spectrum sensing for cognitive radios
has been reported.

There is also a rich literature on MAC protocol design and analysis under
different network and QoS provisioning objectives. In \cite{Su08}, a joint
spectrum sensing and scheduling scheme is proposed where each cognitive user
is assumed to possess two radios. A beacon-based cognitive MAC protocol
is proposed in \cite{Cor07} to mitigate the hidden terminal problem while
effectively exploiting spectrum holes. Synchronized and channel-hopping based
MAC protocols are proposed in \cite{Konda08} and \cite{su08}, respectively.
Other multi-channel MAC protocols \cite{zhang112}, \cite{jha11} are developed 
for cognitive multihop networks. However, these existing papers do not consider
the setting where cognitive radios have access constraints that we investigate
in this paper.

In \cite{tan12}, we have investigated the channel allocation problem
considering this access constraint for a collocated cognitive network
where each cognitive node can hear transmissions from other cognitive nodes (i.e., there is
a single contention domain).
In this paper, we make several fundamental contributions beyond \cite{tan12}.
First, we consider the large-scale cognitive ad hoc network setting in this paper where
there can be many contention domains. In addition, the conflict constraints become
much more complicated since each secondary node may conflict with several
neighboring primary nodes and vice versa. These complex constraints indeed make
the channel assignment and the throughput analysis very difficult. Second, we 
consider a fair channel allocation problem under the max-min fairness criterion \cite{tass05}
while throughput maximization is investigated in \cite{tan12}. Third, we propose
optimal brute-force search and low-complexity channel assignment algorithms and
analyze their complexity. Finally, we develop a
throughput analytical model, which is used in the proposed channel allocation algorithms
and for performance analysis. 


 
\section{System Model and Problem Formulation}
\label{SystemModel}


\subsection{System Model}
\label{System}

We consider a cognitive ad-hoc network where there are $M_s$ flows exploiting spectrum opportunities in  $N$ channels 
for their transmissions. Each secondary flow corresponds to one cognitive transmitter and receiver and we refer to
secondary flows as secondary users (SU) in the following. We assume there are $M_p$ primary users (PU) each of which
can transmit their own data on these $N$ channels.   
We assume that each SU can use at most one channel for his/her data transmission.
In addition, time is divided fixed-size cycle where SUs perform sensing on assigned channels at the beginning of each cycle to 
explore available channels for communications. For simplicity, we assume that there is no sensing error although the analysis
presented in this paper can be extended to consider sensing errors.
It is assumed that SUs transmit at a constant rate which is normalized to 1 for throughput calculation purposes.

\begin{figure}[!t]
\centering
\includegraphics[width=40mm]{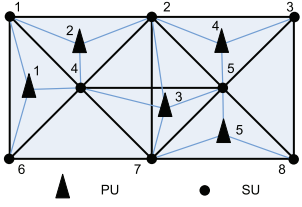}
\caption{The contention graph.}
\label{contgraph}
\end{figure}

To model the interference among SUs in the secondary network, we form a contention graph
$\mathcal{G} = \left\{\mathcal{N},\mathcal{L}\right\}$, where $\mathcal{N}=\left\{1,2,\ldots, M_s\right\}$ is the set of nodes (SUs) representing
SUs and the set of links $\mathcal{L}=\left\{1,2,\ldots, L\right\}$ representing contention relationship among SUs. In particular, there is a link between two SUs in $\mathcal{L}$ if these SUs cannot transmit packet data on the same channel at the same time, which is illustrated in Fig. ~\ref{contgraph}. 
To model the activity of PUs on each channel, let us define $p_{ij}^p$ as the probability that PU $i$ does not transmit on channel $j$.
We stack these probabilities and define $\mathbb{P}_i=\left(p_{i1}^p, \ldots, p_{iN}^p\right)$, $i \in \left[1,M_p\right]$, which
captures the activity of PU $i$ on all channels. In addition, let us define $\mathbb{P}^p = \left( \mathbb{P}_1, \ldots,\mathbb{P}_{M_p} \right)$ where
$\mathbb{P}_i$ is the vector representing activities of PU $i$. 

We now model the contention relationship among SUs and between PUs and SUs. Specifically, we assume that
$\mathcal{U}_i^n$ be the set of neighboring SUs that conflict with SU $i$ (i.e., there is a link connecting each SU in
$\mathcal{U}_i^n$ to SU $i$ in the contention graph). Also, assume that 
SU $k$ has a set of neighboring PUs denoted as $\mathcal{U}_k^p$, which is the subset of ${1,\ldots, M_p}$ so that if
any PU in the set $\mathcal{U}_k^p$ transmit on a particular channel then SU $k$ is not allowed to transmit on this channel
to protect the primary transmission. Assuming that the activities of different PUs on any channel are independent then
the probability that channel $j$ is available for SU $k$ indicates can
be written as $p_{kj}=\prod_{i \in \mathcal{U}_k^p}p_{ij}^p$ since channel $j$ is available for SU $k$ if
all conflicting PUs in $\mathcal{U}_k^p$ do not use channel $j$.

\subsection{Problem Formulation}
\label{ProbForm}

We are interested in performing channel assignment that 
maximizes the minimum throughput among all SUs (i.e., max-min fairness \cite{tass05}).
Let $T_i$ denote the throughput achieved by SU $i$. 
Let $x_{ij}$ describe the channel assignment decision where $x_{ij}=1$ if channel $j$ is assigned to SU $i$ and $x_{ij}=0$, otherwise. 
Then, the max-min channel assignment problem can be written as
\vspace{0.0cm}
\beqn
\label{Tput}
&& \mathop {\max}_\textbf{x} \mathop {\min}_i {T_i}  \label{obj1} 
\eeqn
where $\textbf{x}$ is the channel assignment vector whose elements are $x_{ij}$.
For the case where each SU is allocated a distinct set of channels, i.e., we have 
$ \sum\limits_{i = 1}^{M_s} {{x_{ij}} = 1}, \:  \mbox{for\:all}\: j  \label{con1}$.
Under this non-overlapping channel assignments, let $S_i$ be the set of channels assigned to SU $i$. 
Recall that $p_{ij}$ is the probability that 
channel $j$ is available at SU $i$. Then, $T_i$ can be calculated as 
$T_i = 1 - \prod_{j \in \mathcal{S}_i} \overline{p}_{ij} = 1 - \prod\limits_{j = 1}^N {{{\left( {{{\bar p}_{ij}}} \right)}^{{x_{ij}}}}}$
where $\overline{p}_{ij} = 1 - {p}_{ij}$ is the probability that channel $j$ is not available for SU $i$ \cite{tan12}. 
In fact, $1 - \prod_{j \in \mathcal{S}_i} \overline{p}_{ij}$ is the probability that there is at least one channel available for SU $i$. 
Because each SU can use at most one available channel, its maximum throughput is 1. 

In general, it would be beneficial if each channel is allocated to several SUs in a common 
neighborhood to exploit the multi-user diversity.
Under both non-overlapping and overlapping channel assignments, it can be observed that the channel assignment
problem with the objective defined in (\ref{obj1})
is a non-linear integer program, which is an NP-hard problem (interest readers can refer to \cite{Lee12} for 
detailed treatment of this hardness result). 

\subsection{Optimal Algorithm and Its Complexity}
\label{optalg}

We describe a brute-force search (i.e., exhaustive search) to 
determine the optimal channel assignment solution. Specifically, we can enumerate all possible 
channel assignment solutions then determine the best one by comparing their achieved throughput. While throughput can be calculated
quite easily for the non-overlapping channel assignments as being presented in Section~\ref{ProbForm},
developing a throughput analytical model for an overlapping channel assignment solution is indeed challenging task,
which is performed in Section \ref{tputana} of this paper.

We now quantify the complexity of the optimal brute-force search algorithm.
Let us consider SU $i$ (i.e., $i \in \left\{1, \ldots, M_s\right\}$). Suppose we assign it $k$ channels where $k \in \left\{1, \ldots, N\right\}$). Then, there are $C_N^k$ ways to do so. Since $k$ can take any values in $k \in \left\{1, \ldots, N\right\}$, the total number of ways
to assign channels to SU $i$ is $\sum \limits_{k = 1}^N C_N^k \approx 2^N$. Hence, the total number of ways to assign channels to all SUs is $\left(2^N\right)^{M_s} = 2^{NM_s}$. Recall that 
we need to calculate the throughputs achieved by $M_s$ SUs for each potential assignment to determine the best one.  Therefore, the complexity of the optimal brute-force search algorithm is $\mathcal{O}(2^{NM_s})$.
Given the exponentially large complexity of this brute-force search, we will develop 
low-complexity channel assignment algorithms, namely non-overlapping and overlapping assignment algorithms.

\section{Channel Allocation and Access Design}
\label{nonover}

\subsection{Non-overlapping Channel Assignment}

We develop a low-complexity algorithm for non-overlapping channel assignment in this section.  
Recall that $\mathcal{S}_i$ is the set of channels assigned for secondary
user $i$. In the non-overlapping channel assignment scheme, we have $\mathcal{S}_i \cap \mathcal{S}_j = \emptyset, \: i \neq j$
where SUs $i$ and $j$ are neighbors of each other (i.e., there is a link connecting them in the contention
graph $\mathcal{G}$). Note that one particular channel can be assigned to SUs who are not neighbors of each other.
This aspect makes the channel assignment different from the collocated network setting considered in \cite{tan12}. Specifically
all channels assigned for different SUs should be different in \cite{tan12} under non-overlapping channel assignment
since there is only one contention domain for the collocated network investigated in \cite{tan12}.

The greedy channel assignment algorithm iteratively allocates channels to one of the minimum-throughput SUs so that
we can achieve maximum increase in the throughput for the chosen SU. Detailed description of the proposed algorithm is presented in Algorithm 1.
In each channel allocation iteration, each minimum-throughput SU $i$ calculates its increase in throughput
if the best available channel (i.e., channel  $j_i^* = \mathop {\arg \max }\limits_{j \in {\mathcal{S}_a}} \: {p_{ij}}$)
is allocated. This increase in throughput can be calculated as 
$ \Delta {T_i} = T_i^a - T_i^b = {p_{ij_i^*}} \prod\limits_{j \in \mathcal{S}_i } {(1 - {p_{ij}})}$ \cite{tan12}.

\begin{algorithm}[h]
\caption{\textsc{Non-Overlapping Channel Assignment}}
\label{mainalg}
\begin{algorithmic}[1]

\STATE Initialize SU $i$'s set of available channels, ${\mathcal{S}_i^a} := \left\{ {1,2, \ldots ,N} \right\}$ and $\mathcal{S}_i := \emptyset$ for $i=1, 2,\ldots , M_s$ where $\mathcal{S}_i$ denotes the set of channels assigned for SU $i$.

\STATE $\text{continue} := 1$

\WHILE {$\text{continue} = 1$}

\STATE Find the set of SUs who currently achieve the minimum throughput $\mathcal{S}^{\text{min}} = \mathop {\argmin } \limits_i T_i^b $ where $\mathcal{S}^{\text{min}} = \left\{i_1, \ldots , i_m\right\} \subset \left\{1, \ldots, M_s\right\}$ is the set of minimum-throughput SUs.

\IF {$\mathop {\mathcal{OR}} \limits_{i_l \in \mathcal{S}^{\text{min}}} \left(\mathcal{S}_{i_l}^a \neq \emptyset\right)$}

\STATE For each SU $i_l \in \mathcal{S}^{\text{min}}$ and channel $j_{i_l} \in \mathcal{S}_{i_l}^a$, find $\Delta T_{i_l} = T_{i_l}^a - T_{i_l}^b$ where $T_{i_l}^a$ and $T_{i_l}^b$ are the throughputs after and before assigning channel $j_{i_l}$; and we set $\Delta T_{i_l} = 0$ if $\mathcal{S}_{i_l}^a = \emptyset$

\STATE $\left\{i_l^*, j_{i_l^*}^* \right\}= \mathop {\argmax }\limits_{i_l \in \mathcal{S}^{\text{min}}, j_{i_l} \in {\mathcal{S}_{i_l}^a}} \: \Delta {T_{i_l}} \left(j_{i_l}\right)$

\STATE Assign channel $j_{i_l^*}^*$ to SU $i_l^*$.

\STATE Update $\mathcal{S}_{i_l^*} = \mathcal{S}_{i_l^*} \cup j_{i_l^*}^*$ and $\mathcal{S}_k^a = \mathcal{S}_k^a \backslash j_{i_l^*}^*$ for all $k \in \mathcal{U}_{i_l^*}^n$.

\ELSE

\STATE Set $\text{continue} := 0$

\ENDIF

\ENDWHILE

\end{algorithmic}
\end{algorithm}

In step 4, there may be several SUs achieving the minimum throughput. We denote this set of minimum-throughput SUs as $\mathcal{S}^{\text{min}}$.
Then, we assign the best channel that results in the maximum increase of throughput among all SUs in the set $\mathcal{S}^{\text{min}}$. We
update the set of available channels for each SU after each allocation. Note that only neighboring SUs compete for the same channel; hence,
the update of available channels for the chosen minimum-throughput SU is only performed for its neighbors. This means that we can exploit
spatial reuse in a large cognitive ad hoc network. It can be
verified that if the number of channels is sufficiently large (i.e.,  $N \!\!>>\!\! \max_i |\mathcal{U}_i^n|$), then the proposed non-overlapping channel assignment achieves throughput close to 1 for all SUs.

\subsection{Overlapping Channel Assignment}
\label{over}

\subsubsection{MAC Protocol}

Overlapping channel assignment can improve the minimum throughput but we need to design a MAC protocol
to resolve access contention among different SUs. Note that a channel assignment solution needs to be
determined only once while the MAC protocol operates repeatedly using the chosen channel assignment solution in each cycle. 
Let $\mathcal{S}_i$ be the set of
channels  solely assigned  for SU $i$ and $\mathcal{S}_{i}^{\sf com}$ be the set of  channels assigned for SU $i$ and some other SUs. 
These two sets are referred to as \textit{separate set} and \textit{common set} in the following. 
Let denote $\mathcal{S}_i^{\sf tot} = \mathcal{S}_i \cup \mathcal{S}_i^{\sf com}$, which is the set of all channels assigned to SU $i$. 

Assume that there is one control channel, which is always available and used for access contention resolution.
We consider the following MAC protocol run by any particular SU $i$, which belongs the class of synchronized MAC
protocol \cite{mo08}.\footnote{Since we focus on the channel assignment issue in this paper,
we do not attempt different alternative MAC protocol designs. Interest readers can refer to
\cite{mo08} for detailed treatment of this issue.} 
The MAC protocol operates a cyclic manner where
synchronization and sensing phases are employed before the channel contention and transmission phase in each cycle. 
 After sensing the assigned channels in the sensing phase,
if a particular SU $i$ finds at least one channel in $\mathcal{S}_i$ available, then it chooses
one of these available channels randomly for communication. If this is not the case, SU $i$ will choose one available channel
 in $\mathcal{S}_i^{\sf com}$ randomly (if any). 
Then, it chooses a random backoff value which is uniformly distributed in $[0, W-1]$ (i.e.,  $W$ is the contention window) and 
starts decreasing its backoff counter while listening on the control channel. 

If it overhears transmissions of RTS/CTS from any other SUs, it will freeze from decreasing its backoff counter until the control
channel is free again. As soon as a SU's backoff counter reaches zero, its transmitter and receiver exchange RTS/CTS messages
containing the chosen available channel for communication. If the RTS/CTS message exchange fails due to collisions, the 
corresponding SU will quit the contention and wait until the next cycle.
In addition, by overhearing RTS/CTS messages of neighboring SUs, which convey information about the channels chosen for communications, other SUs compared these channels with their chosen ones. Any SU who has his/her chosen channel 
coincides with the overheard channels quits the contention and waits until the next cycle. 
Note that in the considered cognitive ad hoc setting each SU $i$ only competes with its neighbors in the set $\mathcal{U}_i^n$,
which is different from the setting investigated in \cite{tan12}.

\subsubsection{Throughput Analysis}
\label{tputana}

To analyze the throughput achieved by one
particular SU $i$, we consider all possible sensing outcomes for
the considered SU $i$ on its assigned channels. We will consider the following cases.

\begin{itemize}
\item{ Case 1: If there is at least one channel in $\mathcal{S}_i$ available, then SU $i$ will exploit this available
channel and achieve the throughput of one. Here, we have
\beqn
T_i \left\{\text{Case 1} \right\}  = \Pr\left\{\text{Case 1} \right\} = 1-\prod _{j\in \mathcal{S}_i} \bar{p}_{ij}. \nonumber 
\eeqn }


\item $\text{Case 2}$: We consider scenarios where all channels in $\mathcal{S}_i$ are not 
available; there is at least one channel in $\mathcal{S}_i^{\text{ com}}$ available, and SU $i$ chooses the available channel $j$ for transmission.
Suppose that channel $j$ is shared  by SU $i$ and $\mathcal{MS}_j$ neighboring SUs (i.e., $\mathcal{MS}_j = |\mathcal{U}_j|$ where 
$\mathcal{U}_j$ denotes the set of these $\mathcal{MS}_j$ neighboring SUs). Recall that all $\mathcal{MS}_j$ SUs conflict with SU $i$ (i.e., they are not allowed to transmit data on the same channel with SU $i$). 
There are four possible groups of SUs $i_k$, $k=1, \ldots, \mathcal{MS}_j$ sharing channel $j$, which are described in
the following
\begin{itemize}
\item{ \textbf{Group I}: channel $j$ is not available for SU $i_k$. }
\item{ \textbf{Group II}: channel $j$ is available for SU $i_k$ and SU $i_k$ has at least 1 channel in $\mathcal{S}_{i_k}$ available.}
\item{ \textbf{Group III}: channel $j$ is available for SU $i_k$, all channels in $\mathcal{S}_{i_k}$ are not available and there is another channel
 $j'$ in $\mathcal{S}_{i_k}^{\text{com}}$ available for SU $i_k$. In addition, SU $i_k$ chooses channel $j' \neq j$ for transmission in the contention stage.}
\item{ \textbf{Group IV}: channel $j$ is available for SU $i_k$, all channels in $\mathcal{S}_{i_k}$ are not available. Also, SU $i_k$  chooses channel $j$ for transmission in the contention stage. Hence, SU $i_k$ competes with SU $i$ for channel $j$.}
\end{itemize}

\end{itemize}

Let $\mathcal{U}_{j,i}^p$ be the set of PUs who are neighbors of SUs in $\mathcal{U}_{j}$. 
Then, the throughput achieved by SU $i$ can be written as
\beqn
\label{Tputa1delta}
 T_i\left( \text{ Case 3} \right) =  
 (1-\delta) \Theta_i \sum \limits_{A_1 = 0}^{\mathcal{MS}_j} \sum \limits_{A_2 = 0}^{\mathcal{MS}_j-A_1} \sum \limits_{A_3=0}^{\mathcal{MS}_j-A_1-A_2}
 \frac{1}{1+A_4}  \hspace{0cm} \nonumber \\ 
\sum \limits_{c_1 = 1}^{C_{\mathcal{MS}_j}^{A_1}} \sum \limits_{c_2 = 1}^{C_{\mathcal{MS}_j - A_1}^{A_2}} \sum \limits_{c_3 = 1}^{C_{\mathcal{MS}_j-A_1-A_2}^{A_3}} 
 \Theta_j \Phi_1(A_1) \Phi_2(A_2)  \Phi_3(A_3) \nonumber 
\eeqn
where $A_4= \mathcal{MS}_j-A_1-A_2-A_3$ and $\delta$ denotes the MAC protocol overhead, which will be derived in Section~\ref{Overcal}.
In this derivation, we consider all possible cases where SUs in $\mathcal{U}_{j}$ are divided into four groups
defined above with sizes $A_1$, $A_2$, $A_3$, and $A_4$, respectively. For one such particular case,
let $\mathcal{U}_{j,i}^{p,1}$ be the set of PUs who are only neighbors of SUs in group I with size $A_1$ and
$\mathcal{U}_{j,i}^{p,2}= \mathcal{U}_{j,i}^{p} \backslash \mathcal{U}_{j,i}^{p,1}$ be the set of remaining
PUs in $\mathcal{U}_{j,i}^{p}$. In addition, let $\mathcal{U}_{j,i}^{p,3}$ be the set of PUs who are
 neighbors of SUs in group III and IV with sizes $A_3$ and $A_4$, respectively.
The terms $\Theta_i $ , $\Theta_j $, $\Phi_1(A_1)$, $\Phi_2(A_2)$, and
$\Phi_3(A_3)$ in the above derivation are
\begin{itemize}
\item $\Theta_i $ is the probability that all channels in $\mathcal{S}_i$ are not available and SU $i$ chooses 
an available channel $j$ in $\mathcal{S}_i^{\text{com}}$ for transmission.
\item $\Theta_j$ is the probability that all PUs in $\mathcal{U}_{j,i}^{p,2}$ do not use channel $j$. 
\item $\Phi_1(A_1)$ denotes the total probability of all cases for PUs in $\mathcal{U}_{j,i}^{p,1}$
such that channel $j$ is not available for all $A_1$ SUs in group I.
\item $\Phi_2(A_2)$ represents the probability that there is at least one available channel 
in the separate set for each of the $A_2$ SUs in Group II.
\item $\Phi_3(A_3)$ describes the total probability of all cases for PUs in $\mathcal{U}_{j,i}^{p,3}$ such that
each SU in group III chooses other available channel $j^{'} \neq j$ for transmission and each SU in group IV chooses channel
$j$ for transmission.
\end{itemize}
In this formula, we have considered all possible events and combinations that can happen
for neighboring SUs of the underlying SU $i$. Note that only $A_4$ SUs in Group IV compete
with SU $i$ for channel $j$ by using the proposed MAC protocol. Therefore, SU $i$ wins this contention
with probability $1/(1+A_4)$. In addition,
the throughput is reduced by a factor $1-\delta$ where $\delta$ is the MAC protocol overhead.
Due to the space constraint, detailed derivation of these rather complicated probabilities
are presented in the online technical report.
 Summarizing all considered cases, the throughput achieved by SU $i$ is given as 
\beqn
\label{T_over_final}
T_i = T_i \left\{\text{Case 1} \right\} + T_i \left\{ \text{Case 2} \right\} .
\eeqn
This throughput derivation is used for channel assignment and performance evaluation of the proposed algorithms.

\subsubsection{Configuration of Contention Window}
\label{ConWinCal}

We show how to calculate contention window $W$ so that collision probabilities among contending SUs are sufficiently small.
Note that the probability of the first collision among potential collisions is largest because the number of contending SUs decreases
for successive potential collisions. Derivation of these collision probabilities for the cognitive ad-hoc networks
is more complicated than that for collocated networks considered in \cite{tan12} since
the interference constraints are more complicated. 
 
We calculate contention window $W_k$ for each SU $k$ considering the contention with its neighbors. 
Let us calculate $\mathcal{P}_{c,k}$ as a function of $W_k$ assuming that there are $m$  secondary 
 SUs  in the contention phase. Without loss of generality, assume that the random backoff times of $m$  SUs are ordered as $r_1 \leq r_2 \leq \ldots \leq r_m$. The conditional probability of the first collision if there are $m$  SUs in the contention stage can be written as
\beqn
\label{Pfirstc}
\mathcal{P}_{c,k}^{(m)} &=& \sum _{j=2}^{m} \Pr \left( j \: \text{users collide} \right) \nonumber\\
&=& \sum_{j=2}^m \sum_{l=0}^{W_k-2} C_m^j \left( \frac{1}{W_k}\right)^j \left( \frac{W_k-l-1}{W_k}\right)^{m-j}
\eeqn
where each term in the double-sum represents the probability that $j$ users collide when they
choose the same backoff value equal to $l$. 
Hence, the probability of the first collision can be calculated as 
\beqn \label{pc}
\mathcal{P}_{c,k} = \sum_{m=2}^{M_k^n} \mathcal{P}_{c,k}^{(m)} \times \Pr\left\{m \: \text{ users contend}\right\},
\eeqn
where $M_k^n = \left|\mathcal{U}_k^n\right|+1$ is the total number of SUs (including SU $k$ and its neighbors), $\mathcal{P}_{c,k}^{(m)}$ is given in (\ref{Pfirstc}) and $\Pr\left\{m  \: \text{users contend}\right\}$
is the probability that $m$ SUs contend with SU $k$ in the contention phase. To compute $\mathcal{P}_{c,k}$, we now
derive $\Pr\left\{m  \: \text{users contend}\right\}$. 

We can divide the set of neighbors of SU $k$ into two groups. In particular, there are $m$ SUs contending with SU $k$ while the remaining $M_k^n - m$ SUs do not join the contention phase. There are $C_{M_k^n}^m$ such combinations for a particular value of $m$ where it happens with the
following probability
\beqn \label{Pmusercon}
\Pr \left\{ m \: \text{users contend} \right\} = \sum_{n=1}^{C_{M_k^n}^m} \mathcal{P}_{\text{con}}^{(n)} 
\eeqn
where $\mathcal{P}_{\text{con}}^{(n)}$ is the probability of one particular case where $m$ SUs contend with SU $k$.
We can divide the set of remaining $M_k^n - m$ SUs who do not join the contention into two subgroups, namely
SUs who could not find any available channels in their allocated channels $\mathcal{S}_{i_2}^{\sf tot}$ (first subgroup) and
SUs who find some available channels in their separate sets $\mathcal{S}_{i_1}$ (second subgroup).

Now, let ${\Lambda}_n$ be one particular set of $m$ SUs in the first group and $A_1$ denote the number of SUs in
the first subgroup of the remaining $M_k^n - m$ SUs. Then, we can calculate $\mathcal{P}_{\text{con}}^{(n)}$
as follows:
\beqn
\mathcal{P}_{\text{con}}^{(n)} = \prod \limits_{i_1 \in \Lambda_n} \left[ \prod \limits_{l_1 \in \mathcal{S}_{i_1}} \overline{p}_{i_1l_1} \right] \hspace{4.1cm} \label{1term}  \\
\sum \limits_{A_1=0}^{M_k^n-m} \sum \limits_{c_1=1}^{C_{M_k^n-m}^{A_1}} \prod \limits_{i_2 \in \Omega_{c_1}^{(1)}} \prod \limits_{l_2 \in \mathcal{S}_{i_2}} \overline{p}_{i_2l_2}  \prod \limits_{i_3 \in \Omega_{c_1}^{(2)}} \left(1- \prod \limits_{l_3 \in \mathcal{S}_{i_3}} \overline{p}_{i_3l_3}\right) \label{2term}  
\eeqn
\beqn
\sum \limits_{n^{(1)} =1}^{\beta^{(1)}} \sum \limits_{q^{(1)} =1}^{C_{\beta^{(1)}}^{n^{(1)}}}  \ldots \sum \limits_{n^{(m)} =1}^{\beta^{(m)}} \sum \limits_{q^{(m)} =1}^{C_{\beta^{(m)}}^{n^{(m)}}}   \prod_{i_4 \in \mathcal{U}_{c_1}^p } \prod_{l_4 \in \Lambda_{c_1}^{(1)}} p^p_{i_4l_4} \prod_{l_5 \in \Lambda_{c_1}^{(2)}}  \overline{p}^p_{i_4l_5}.   \label{3term} 
\eeqn
The term inside $[.]$ in (\ref{1term}) represents the probability that 
all channels in the separate sets $\mathcal{S}_{i_1}$ for all SUs $i_1 \in \Lambda_n$ are not available
so that these SUs contend to access available channels in $\mathcal{S}_{i_1}^{\sf com}$. The term in (\ref{2term})
denotes the probability that each of $A_1$ SUs in the first subgroup (i.e., in the set $\Omega_{c_1}^{(1)}$) find no
available channels in their separate sets and each of the $M_k^n-m-A_1$ SUs in the second subgroup (i.e., in the set $\Omega_{c_1}^{(2)}$)
find at least one  available  channel in their separate sets (therefore, these SUs will not perform contention). 
Here, $c_1$ is the index of one particular case where there are $A_1$ SUs in the first subgroup and a particular set ${\Lambda}_n$.
The last term in (\ref{3term}) denotes the probability of the event representing the status of all PUs who are neighbors of
SUs in the set $\mathcal{U}_k^n$ (i.e., neighbors of SU $k$) so that there are exactly $m$ contending SUs in the set ${\Lambda}_n$
and  $A_1$ SUs in the first subgroup. In (\ref{3term}) we consider all possible
scenarios where for each SU $i \in \Lambda_n$, there are $n^{(i)}$ available channels among $\beta^{(i)} = |\mathcal{S}_{i}^{\sf com}|$ 
channels in the set $\mathcal{S}_{i}^{\sf com}$ where
$q^{(i)}$ represents the index of one such particular case. Corresponding to such $(n^{(i)},q^{(i)})$, $\mathcal{U}_{c_1}^p$ denotes
the set of PUs who are neighbors of SUs in $\mathcal{U}_k^n$ so that indeed $m$ underlying SUs perform contention.

By substituting $\mathcal{P}_{\text{con}}^{(n)}$ calculated above into (\ref{Pmusercon}), we can calculate
the collision probability in $\mathcal{P}_{c,k}$ in (\ref{pc}). From this, we can determine $W_k$ as follows:
\beqn \label{Window}
W_k=  \min \left\{{W_k} \: \text{such that} \:  \mathcal{P}_{c,k}(W_k) \leq \epsilon_{P_k} \right\}
\eeqn
where $\epsilon_{P_k} $ controls the collision probability and overhead tradeoff and
 for clarity we denote $\mathcal{P}_{c,k}(W_k)$, which is given in (\ref{pc}) as a function of $W_k$.
Then, we will determine the contention window for all SUs as $W = \max_k W_k$.

\subsubsection{Calculation of MAC Protocol Overhead}
\label{Overcal}

Let $r$ be the average value of the backoff value chosen by any SU.
Then, we have $r = (W-1)/2$ because the backoff counter value is uniformly chosen
in the interval $[0,W-1]$. As a result, average overhead can be calculated as follows:
\beqn \label{overhead}
\delta\left(W\right) = \frac { \left[ W-1 \right]\theta/2 + t_{\sf RTS} + t_{\sf CTS} + 3 t_{\sf SIFS} + t_{\sf SEN}+ t_{\sf SYN}} {\sf T_{\sf cycle}} 
\nonumber
\eeqn
where $\theta$ is the time corresponding to one backoff unit;  $t_{\sf RTS}$,  $t_{\sf CTS}$, $t_{\sf SIFS}$
are the corresponding time of RTS, CTS and SIFS (i.e., short inter-frame space) messages; $t_{\sf SEN}$ is the
sensing time; $t_{\sf SYN}$ is the transmission time of the synchronization message; 
and $\sf T_{\sf cycle}$ is the cycle time. 

\subsubsection{Overlapping Channel Assignment Algorithm}

\begin{algorithm}[h]
\caption{\textsc{Overlapping Channel Assignment}}
\label{mainalg}
\begin{algorithmic}[1]


\STATE After running Algorithm 1, each SU $i$ has $\mathcal{S}_i$, $\mathcal{S}_i^{\text{com}}=\emptyset$ and $\mathcal{S}_i^n$, $i=1,\ldots,M_s $.

\STATE continue := 1.

\WHILE {$\text{continue}  = 1$}

\STATE Find $T_{\text{min}}$ and $i^* = \mathop {\argmin} \limits_{i \in \left\{1,\ldots, M_s \right\}} T_i^b$.



\STATE $\mathcal{S}_{i^*}^{\text{Uni}} = \mathop \bigcup \limits_{l \in \mathcal{U}_{i^*}^n } \mathcal{S}_l^{\text{tot}}  $.

\STATE $\mathcal{S}_{i^*}^{\text{Sep}} = \mathop {\text{SETXOR}} \limits_{l \in \mathcal{U}_{i^*}^n } \left(\mathcal{S}_l^{\text{tot}} \right) $.

\STATE $\mathcal{S}_{i^*}^{\text{Int}} = \mathcal{S}_{i^*}^{\text{Uni}} \backslash \mathcal{S}_{i^*}^{\text{Sep}} \backslash \mathcal{S}_{i^*}^{\text{com}}$.

\STATE Find all minimum-throughput SUs and find the best channels from either $\mathcal{S}_{i^*}^{\text{Sep}}$ or 
$\mathcal{S}_{i^*}^{\text{Int}}$ for these minimum-throughput SUs to improve the overall minimum throughput.

\IF {$\mathop \bigcup \limits_{i \in \left\{1,\ldots, M_s \right\}} \mathcal{S}_i^{\text{com},\text{temp}} \neq \emptyset$}

\STATE Assign $\mathcal{S}_i^{\text{com}} = \mathcal{S}_i^{\text{com},\text{temp}}$ and $\mathcal{S}_i = \mathcal{S}_i^{\text{temp}}$.

\ELSE

\STATE Set $\text{continue} := 0$.

\ENDIF

\ENDWHILE

\end{algorithmic}
\end{algorithm}


In the overlapping channel assignment algorithm described in Algorithm 2, we  run Algorithm 1 to
 obtain the non-overlapping channel assignment solution in the first phase and perform overlapping channel assignments
by allocating channels that have been assigned to a particular SU
to other SUs in the second phase. 
We calculate the increase-of-throughput metric for all potential
channel assignments that can improve the throughput of minimum-throughput SUs.
To calculate the increase-of-throughput, we use the throughput analytical model
in Subsection \ref{tputana}, where the MAC protocol overhead, $\delta<1$ is derived from \ref{Overcal}.
After running Algorithm 1 in the first phase, each SU $i$ has the set of assigned non-overlapping channels, $\mathcal{S}_i$, and 
it initiates the set of overlapping channels as $\mathcal{S}_i^{\text{com}}=\emptyset$, $i=1,\ldots,M_s $. 
Recall that the set of all assigned channels for SU $i$ is $\mathcal{S}_{i}^{\text{tot}}=\mathcal{S}_{i} \cup \mathcal{S}_{i}^{\text{com}}$. 
Let $\mathcal{S}_{i^*}^{\text{Uni}}$ is the set of all channels that have been assigned for SU $i^*$'s neighboring SUs.
Also, let $\mathcal{S}_{i^*}^{\text{Sep}}$ be the set of all channels assigned solely for each individual neighbor of SU $i^*$ (i.e., each
channel in $\mathcal{S}_{i^*}^{\text{Sep}}$ is allocated for only one particular SU in $\mathcal{U}_{i^*}^n$).
Therefore, $\mathcal{S}_{i^*}^{\text{Int}} $ defined in step 7 of Algorithm 2 is the set of ``intersecting channels'', which are shared by 
at least two neighbors of SU $i^*$. Here, SETXOR($\textbf{A,B}$) would return the set of all elements in $\textbf{A}$ or $\textbf{B}$ but not the common elements of both $\textbf{A}$ and $\textbf{B}$. 

In each iteration, we determine the set of SUs which achieve the minimum throughput.
Then, we need to search over two sets $\mathcal{S}_{i^*}^{\text{Sep}}$ or 
$\mathcal{S}_{i^*}^{\text{Int}}$ to find the best channel for each of these minimum-throughput SUs. Note
that allocation of channels in $\mathcal{S}_{i^*}^{\text{Int}}$ to minimum-throughput SUs can indeed decrease
the achievable throughput of their owners (i.e., SUs which own these channels before the allocation).
Therefore, channel allocations in step 8 are only performed if the minimum throughput can be improved.
In step 9, $\mathcal{S}_i^{\text{com},\text{temp}}$ is the potential set of channels for SU $i$.
Algorithm 2 terminates when there is no assignment that can improve the minimum throughput.
Due to the space constraint, detailed description of step 8 is omitted.

\subsubsection{Complexity Analysis}
\label{Compx}
In each iteration of Algorithm 1, the number of minimum-throughput SUs is at most $M_s$
and there are at most $N$ channel candidates which can be allocated for each of them.
Therefore, the complexity involved in each iteration is upper bounded by $M_s N$.
We can also determine an upper bound for the number of iterations, which is $M_s N$.
This is simple because each SU can be allocated at most $N$ channels and there are $M_s$ SUs.
Therefore, the complexity of Algorithm 1 is upper bounded by $M_s^2 N^2$. In Algorithm 2,
we run Algorithm 1 in the first phase and perform overlapping channel assignments in the second
phase. The complexity of this second phase can also be upper-bounded by $M_s^2 N^2$.
Therefore, the complexity of both Algorithms 1 and 2 can be upper-bounded by $\mathcal{O} \left(M_s^2 N^2 \right)$,
which is much lower than that of the brute-force search algorithm presented in Section~\ref{optalg}.

\section{Numerical Results}
\label{Results}

To obtain numerical results, we choose the length of control packets as follows: RTS including PHY header 288 bits, CTS including PHY header 240 bits, which correspond to $t_{\text{RTS}} = 48\mu s$, $t_{\text{CTS}} = 40\mu s$ for transmission rate of 6 Mbps, which is the basic rate of 802.11a/g standards \cite{802S}. Other parameters are chosen as follows: cycle time $T_{\text{cycle}} = 3ms$; $\theta = 20 \mu s$, $t_{\text{SIFS}} = 28 \mu s$, target collision probability $\epsilon_P = 0.03$; $t_{\text{SEN}} $ and $t_{\text{SYN}}$ are assumed to be negligible so they are ignored. Note that these values of $\theta$ and $t_{\text{SIFS}}$ are typical (e.g., see \cite{Bi00}). 

\begin{figure}[!t]
\centering
\includegraphics[width=60mm]{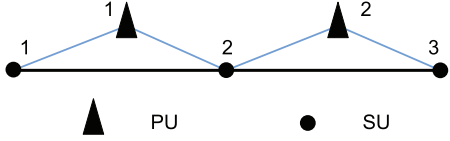}
\caption{The scenario with 3 SUs and 2 PUs.}
\label{contgraph_opt}
\end{figure}

\begin{figure}[!t]
\centering
\mbox{\subfigure[]{\includegraphics[width=1.7in]{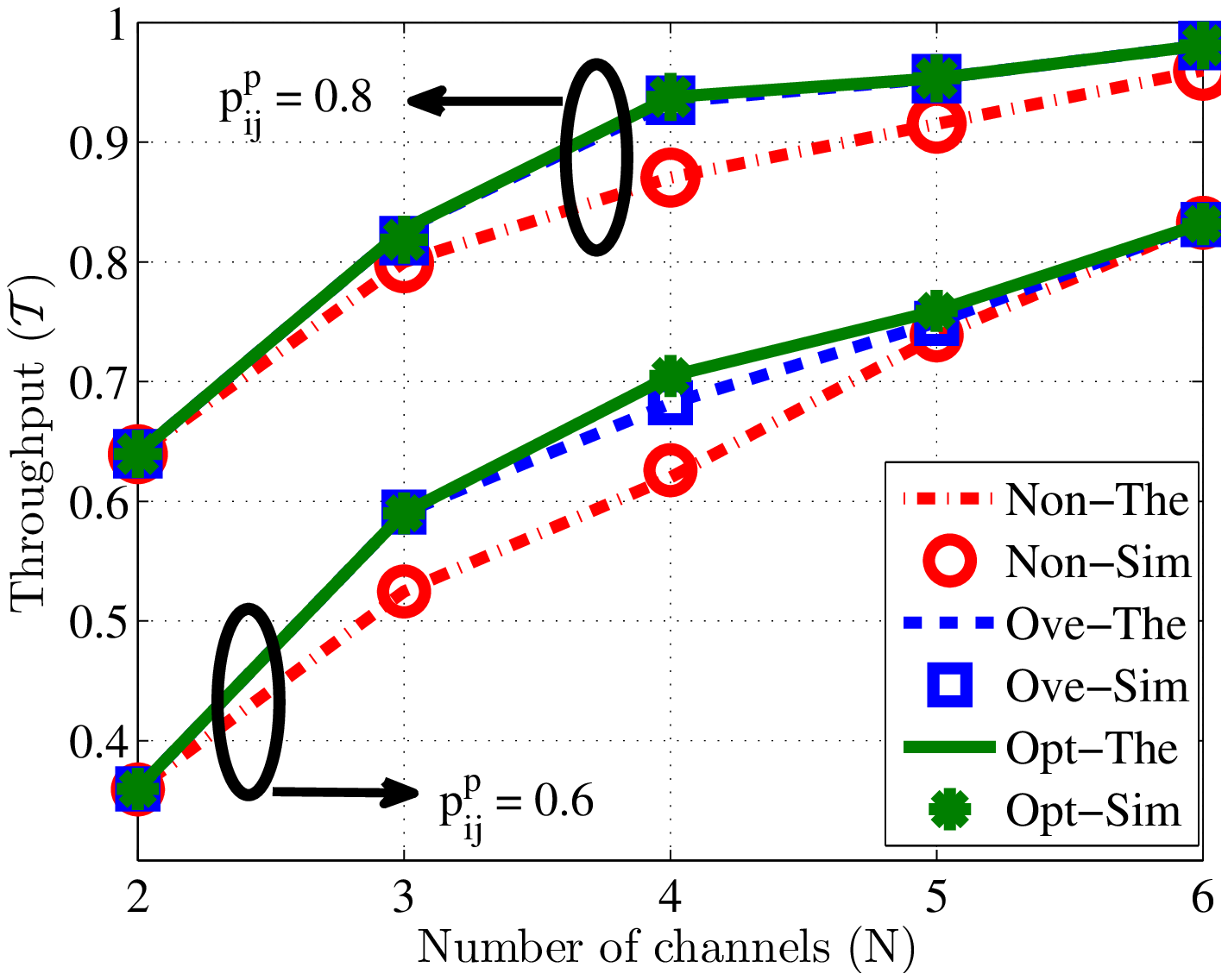} \label{Opt_M3_NP2_Pp06_08}}  
\subfigure[]{\includegraphics[width=1.7in]{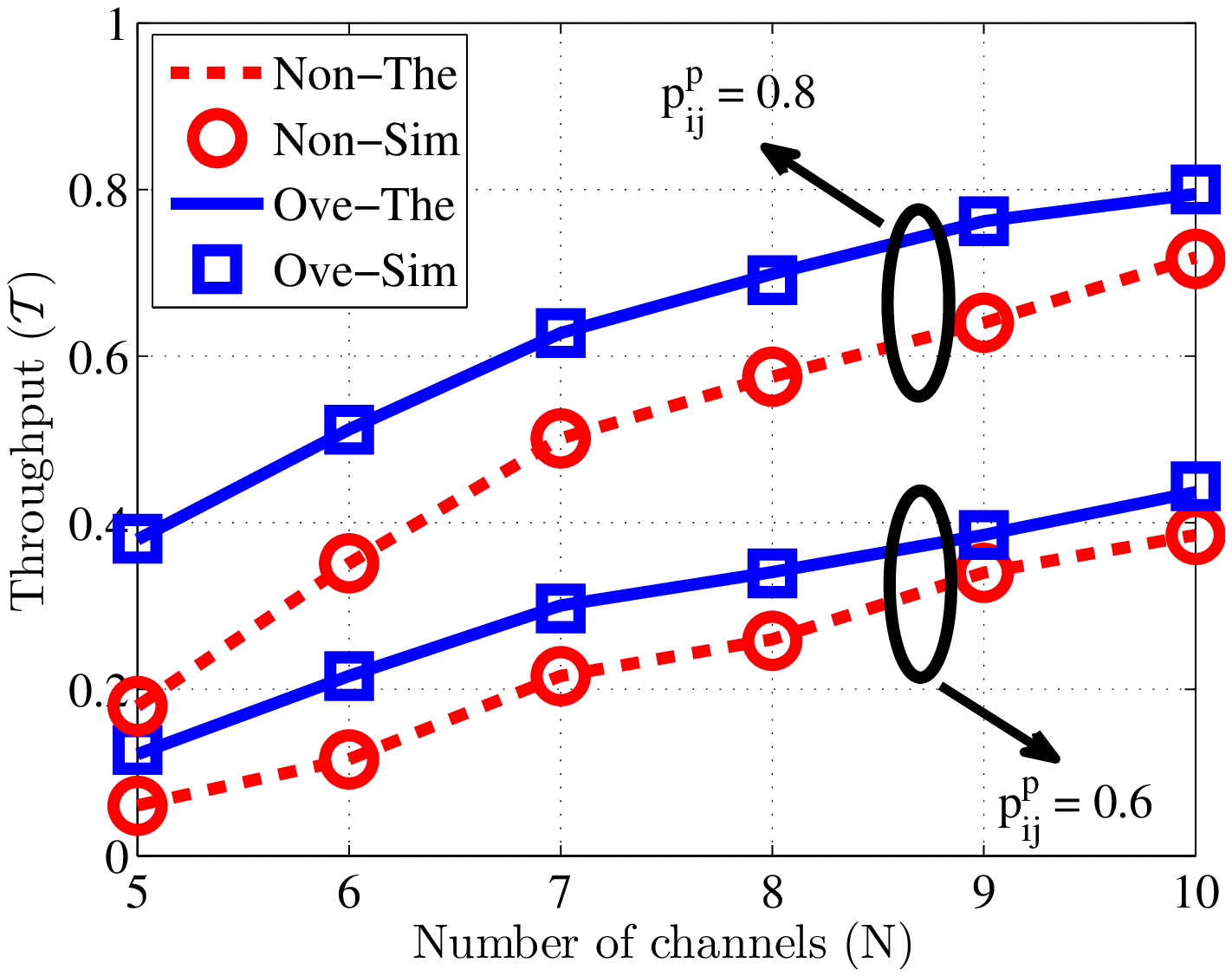} \label{M8_NP5_Pp06_08}} }
\caption{Throughput versus the number of channels, $p_{ij}^p=0.6$ and 0.8, Non: Non-overlapping, Ove: Overlapping, The: Theory, Sim: Simulation, Opt:Optimal.(a) $M_p = 2$, $M_s = 3$ (b) $M_p = 5$, $M_s = 8$}
\end{figure}

\begin{figure}[!t]
\centering
\mbox{\subfigure[]{\includegraphics[width=1.7in]{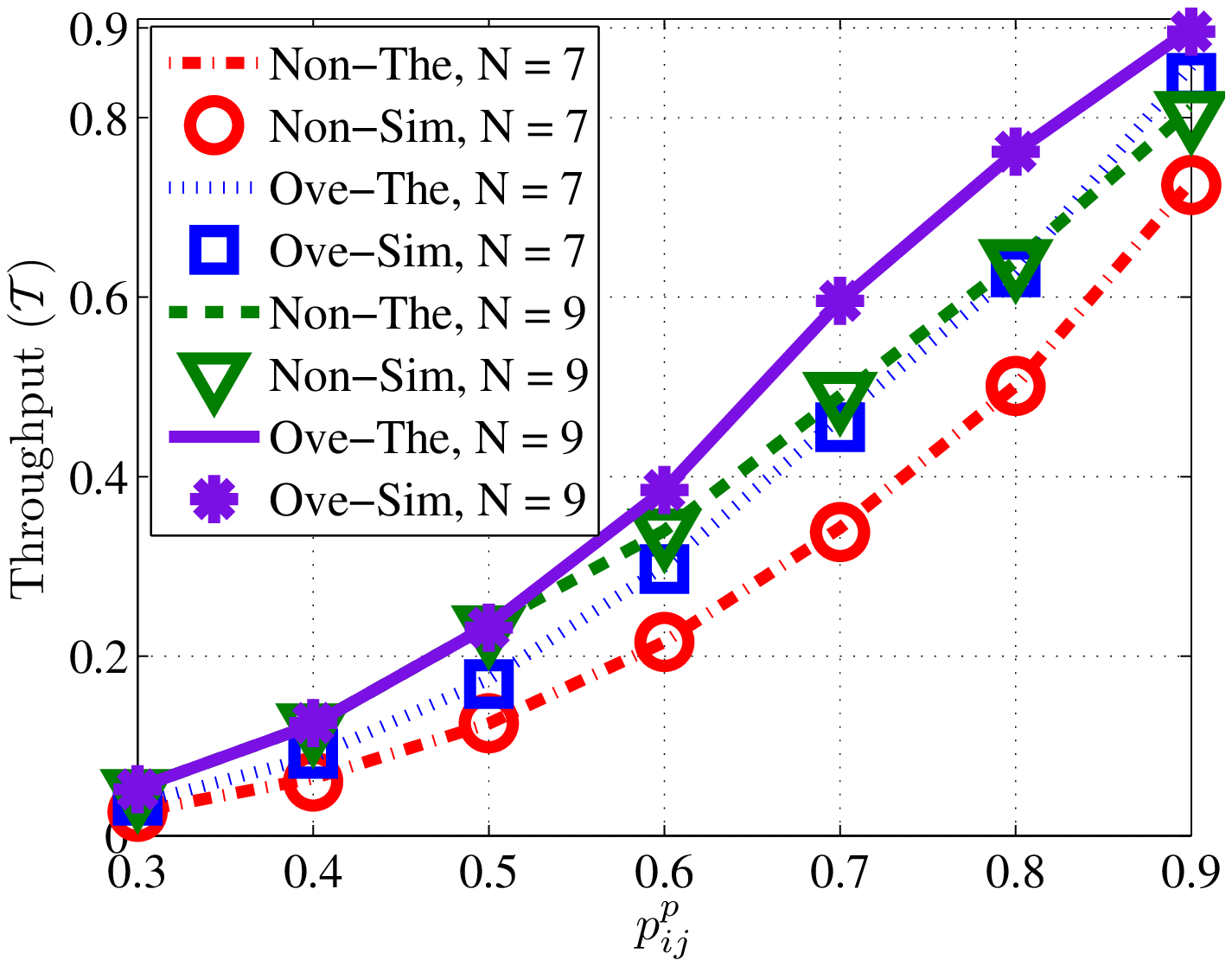} \label{TvsPp_M8_Np5_N7_9}}  
\subfigure[]{\includegraphics[width=1.7in]{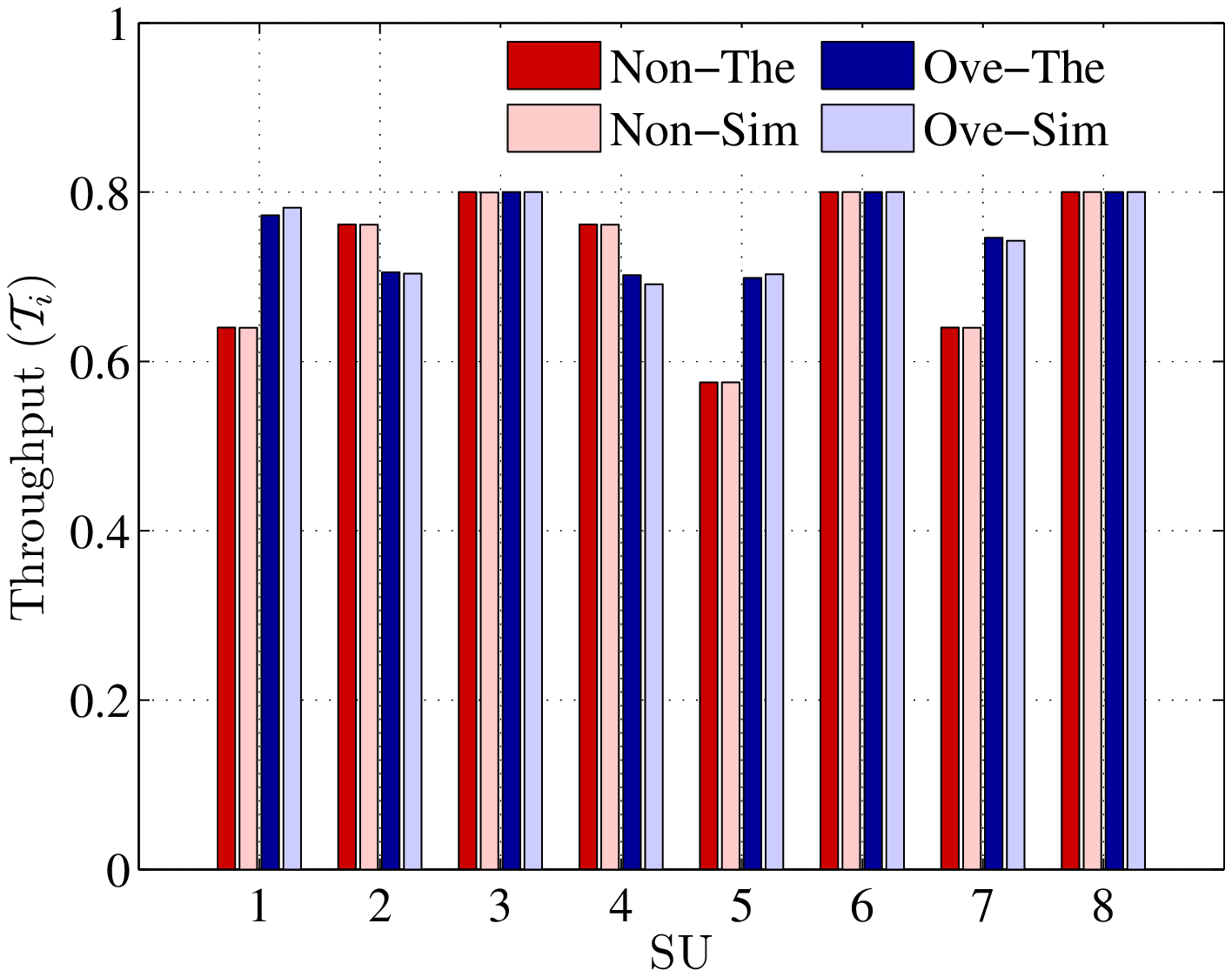} \label{Bar_M8_NP5_Pp08_N8}} }
\caption{(a) Throughput versus $p_{ij}^p$, $N = 7$ and 9. (b) Throughput achieved by each SU, $M_p = 5$, $M_s = 8$, $p_{ij}^p =0.8$, $N$ = 8.}
\end{figure}

To compare the performance of optimal brute-force search and our proposed algorithms, we consider a small network shown in Fig. \ref{contgraph_opt} where we choose $M_s = 3$ SUs, $M_p=2$ PUs and $p_{ij}^p = 0.6$ and 0.8. Fig.~ \ref{Opt_M3_NP2_Pp06_08} shows that the minimum throughputs achieved
by Algs 2 are very close to that obtained the optimal search, which confirms the merit of this low-complexity algorithm.
Also, the simulation results match the analytical results very well, which validates the proposed throughput analytical model.
Figs.~ \ref{M8_NP5_Pp06_08}, \ref{TvsPp_M8_Np5_N7_9}, and \ref{Bar_M8_NP5_Pp08_N8} illustrate the minimum throughputs 
achieved by our proposed algorithms for a larger network shown in Fig.~ \ref{contgraph}. In particular,
Fig.~ \ref{M8_NP5_Pp06_08} shows the minimum throughput versus the number of channels for $p_{ij}^p$ equal to 0.6 and 0.8. 
This figure confirms that Alg. 2 achieves significantly larger throughput than that due to Alg. 1 thanks to overlapping
channel assignments. 
  
Fig.~\ref{TvsPp_M8_Np5_N7_9} illustrates the minimum throughput versus $p_{ij}^p$. It can be observed that as $p_{ij}^p$ increases,
the minimum achievable throughput indeed increases. This figure also shows that the minimum throughput for $N$ = 9 
is greater than that for $N=7$. This means our proposed algorithms can efficiently exploit available
spectrum holes. In Fig.~\ref{Bar_M8_NP5_Pp08_N8}, we illustrate the throughputs achieved by different SUs to demonstrate
the fairness performance. It can be observed that the differences between the maximum and minimum throughputs under
 Alg. 2 are much smaller than that due to Alg. 1. This result implies that Alg. 2 not only achieves better throughput but also
results in improved fairness compared to Alg. 1. 



\vspace{0.2cm}
\section{Conclusion}
\label{conclusion} 

We have investigated the fair channel
allocation problem in cognitive ad hoc networks. Specifically, we have presented both 
optimal brute-force search and low-complexity algorithms and analyzed
their complexity and throughput performance through analytical and numerical studies.

\bibliographystyle{IEEEtran}

\end{document}